\documentclass[10pt,journal]{IEEEtran}

\usepackage[pdftex]{graphicx}
\DeclareGraphicsExtensions{.pdf,.jpeg,.png}

\usepackage{multirow,setspace,verbatim,amsfonts,amsthm,epsfig}
\usepackage{graphicx,amsmath,hyperref, cite}
\usepackage{amsbsy,amssymb,booktabs}
\usepackage{makecell,color}

\usepackage{array}
\usepackage{booktabs}

\newcommand\mdoubleplus{\mathbin{+\mkern-10mu+}}

\begin{document}

\title{E3-targetPred: Prediction of E3-Target Proteins Using Deep Latent Space Encoding}

\author{Seongyong Park, Shujaat Khan, Abdul Wahab
\thanks{Seongyong Park and Shujaat Khan are with the Department
of Bio and Brain Engineering, Korea Advanced Institute of Science and Technology, Dae-jeon, South Korea,
34141 e-mail: \{sypark0215,shujaat\}@kaist.ac.kr, Abdul Wahab is with the Department of Mathematics, School of Natural Sciences, National University of Sciences and Technology (NUST), Sector H-12, 44000, Islamabad, Pakistan. e-mail: abdul.wahab@sns.nust.edu.pk}}

\maketitle

\begin{abstract}
Understanding E3 ligase and target substrate interactions are important for cell biology and therapeutic development. However, experimental identification of E3 target relationships is not an easy task due to the labor-intensive nature of the experiments. In this article, a sequence-based E3-target prediction model is proposed for the first time. The proposed framework utilizes \emph{composition of k-spaced amino acid pairs} (CKSAAP) to learn the relationship between E3 ligases and their target protein. A class separable latent space encoding scheme is also devised that provides a compressed representation of feature space. A thorough ablation study is performed to identify an optimal gap size for CKSAAP and the number of latent variables that can represent the E3-target relationship successfully. The proposed scheme is evaluated on an independent dataset for a variety of standard quantitative measures. In particular, it achieves an average accuracy of $70.63\%$ on an independent dataset. The source code and datasets used in the study are available at the author's GitHub page (\href{https://github.com/psychemistz/E3targetPred}{https://github.com/psychemistz/E3targetPred}).

\end{abstract}


\IEEEpeerreviewmaketitle
\section{Introduction}

\IEEEPARstart{I}n eukaryotic cells, proteolysis is mediated by two major organelles, lysosomes and proteasomes, however, 80-90\% of the intracellular proteolysis is mediated by proteasomes \cite{lee1998proteasome}. Protein degradation by proteasomes is controlled by the ubiquitination, and in this process, the target substrate is covalently conjugated to a chain of $76$ amino acid poly-peptides called ubiquitin through enzyme cascade. Repetition of this reaction produces a poly-ubiquitin chain on the substrate that can be a target of the $26$S proteasome\cite{von2006nuclear}.  There is a known family of enzymes that mediates covalent attachment of ubiquitin to substrates called E1, E2, and E3. E1 is a ubiquitin-activating enzyme that forms a thiol-ester bond at the carboxy-terminal glycine of ubiquitin. Once activated, E2, called a ubiquitin-conjugating enzyme, catalyzes the trans-thiolation reaction between E1 and E2 to form E2-ubiquitin conjugate. Ubiquitin ligase, E3, mediates the transfer of ubiquitin from the E2-ubiquitin conjugate to the target protein, most commonly onto the $\epsilon$-amino group of a lysine residue on the protein substrate \cite{hoeller2009targeting}. Therefore, E3 ligase is a scaffold protein that recognizes E2 and the target protein simultaneously.

The ubiquitination-proteasome system (UPS) plays a central role in various cellular processes such as cell cycle, signal transduction, gene expression, development, and protein folding \cite{kirschner1999intracellular} and it plays a significant role in disease biology such as neurodegenerative diseases \cite{ross2004ubiquitin} and cancer \cite{ge2018integrated}. Since the substrate specificity of ubiquitination is determined by the E3 ligase, many experimental and computational studies have been conducted to discover the relationship between E3 and their target proteins. Various experimental techniques have been developed to elucidate the E3-target interaction such as Global protein stability (GPS) profiling \cite{yen2008identification}, protein microarray \cite{merbl2009large}, phage display \cite{guo2013screening}, and mass spectroscopy \cite{yumimoto2012comprehensive}. Unfortunately, due to the low expression level of the substrates and their inherently weak interactions, it is hard to find the relationship only by experimental techniques. Towards this end, some of the recent studies proposed novel computational approaches \cite{han2012e3net, nguyen2016ubinet, li2017integrated}. E3Net was designed to comprehensively collecting available E3s and their substrate information through text-mining \cite{han2012e3net}. In \cite{nguyen2016ubinet}, UbiNet combined the  E3Net \cite{han2012e3net} with experimentally verified E3-targets and also rendered some manually curated E3-target relations to provide an integrated resource for E3-target relations.  Recently, Li \emph{et al.} \cite{li2017integrated} designed  the Naive-Bayes model that predicts possible E3-target relationships by combining ortholog, network topology, domain, and function information. 

In this paper, a novel sequence-based E3-target prediction method is proposed for the first time that does not require complicated feature engineering, e.g., extraction of ortholog, network topology, domain, and function information. In particular, our proposed method only uses E3 and target protein sequences to extract \emph{composition of k-spaced amino acid pairs} (CKSAAP). It is hypothesized that there is sufficient discriminatory information available in sequence-derived features that can be learned from the known E3-target relationship to design a generalized E3-Target prediction model. To avoid sequence homology problems, the framework is substantiated through known human E3-target relationships. 

The article is organized as follows. In Section \ref{sec:method}, details of dataset, feature extraction, and latent space encoding are discussed, followed by ablation study and experimental results in Section \ref{sec:results}. In Section \ref{sec:discussion}, a discussion on latent space encoding of predicted E3-target is given. The findings of the article are concluded  in Section \ref{sec:conclusion}.

\section{Proposed Methods} \label{sec:method} 

\subsection{Dataset} \label{sec:dataset} 

The gold standard positive E3-target relationship proposed in literature is used as positive data. In particular, we used $1325$ manually curated E3-target interactions \cite{li2017integrated},  extracted using E3miner (E3 Target Relational Text Mining Tool) for PubMed abstracts. For a gold standard negative dataset, we collected $727$ non-interacting E3 PPIs from Negatome 2.0 \cite{blohm2014negatome} and NIP \cite{zhang2018predicting}.

To design the proposed E3-targetPred model, we randomly selected $600$ positive and $600$ negative E3-target relationships as the training dataset out of which $30$ positive and $30$ negative relationships are randomly selected as the validation set. The remaining samples are used as test dataset. Furthermore, an independent dataset consisting of $1145$ E3 and target relations are collected from ESI network study\cite{chen2019multidimensional} and Negatome 2.0 \cite{blohm2014negatome}.  The independent dataset is designed to test model generalization power for different scenarios, for instance, different seen and unseen E3-Target pairs.  Here seen means E3 or targets which are available in development dataset (that is used for training, validation, and testing case) but their specific pairs are not the part of the development dataset. Similarly, unseen means that the particular E3s or targets are not part of the development dataset.  The statistics of independent test dataset is summarized in Table~\ref{tab:ind_sample_stat}.

\begin{table}[!htb]
\caption{Sample statistics summary of independent test dataset}
\begin{center}
\resizebox{0.45\textwidth}{!}{
\begin{tabular}{c c c c c}
\\ \hline
Dataset & $N_{\rm E3}$ & $N_{\rm Target}$ & $N_{\rm Relation}$ & $N_{\rm Positive}$ \\ \hline\hline 
Seen E3, Seen Target & 108 & 590 & 914 & 321 \\ \hline 
Seen E3, UnSeen Target & 115 & 512 & 940 & 680 \\ \hline 
UnSeen E3, Seen Target & 112 & 250 & 369 & 84 \\ \hline 
UnSeen E3, Unseen Target & 95 & 350 & 415 & 60  \\ \Xhline{2\arrayrulewidth}
Total & 282 & 1,352 & 2,638 & 1,145 \\ \hline
\end{tabular}
}
\end{center}
\label{tab:ind_sample_stat}
\end{table}

\subsection{Features Extraction}\label{features}

\begin{figure*}[!htb]
    \centering 
    \includegraphics [width=18cm]{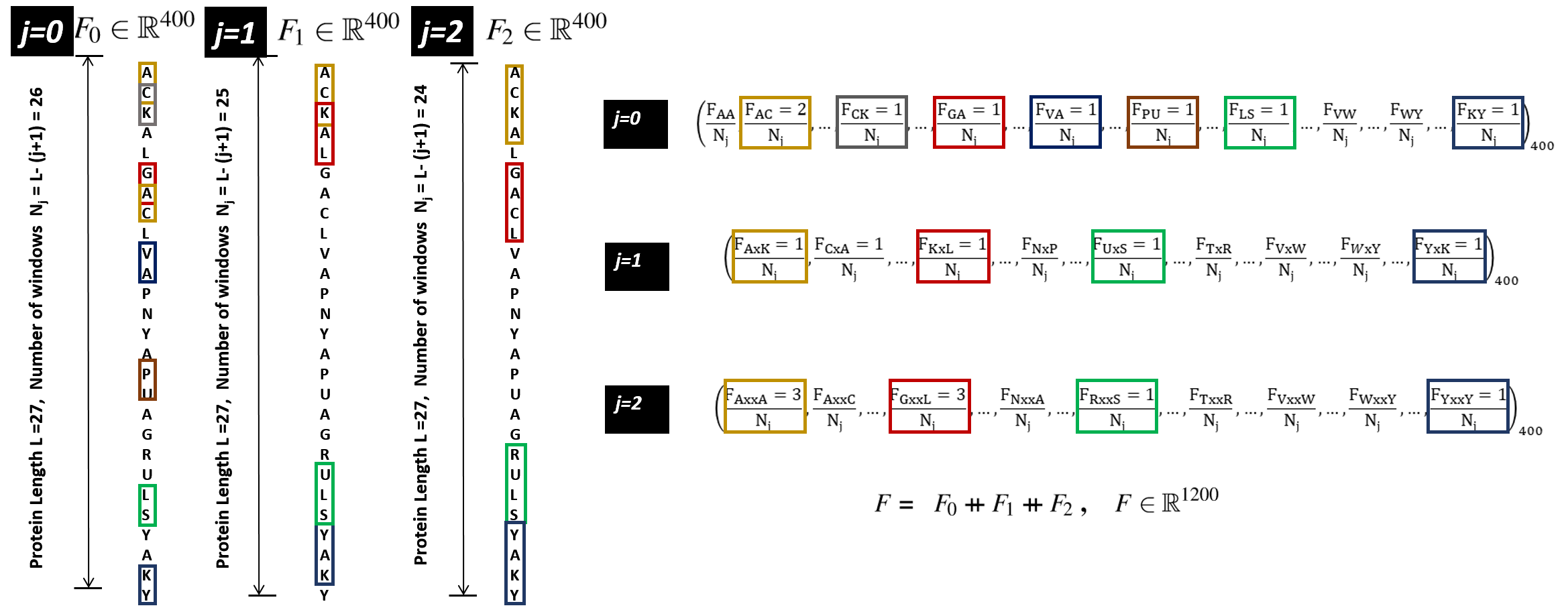}
    \caption{Illustration of CKSAAP descriptor calculation for $k=2$. }
    \label{fig_comp}
\end{figure*}

In order to design a machine learning-based classification model for E3-target prediction, we propose to use the CKSAAP features.  In CKSAAP, frequency of amino-acid pairs is calculated and the concept of pair is defined by a gap of $j=0 \dots k$, i.e., k-spaced amino acid pairs. For example, for $k=2$, three different forms of pairs are calculated that are spaced by zero, one and two peptide residues. An illustrative example of CKSAAP composition is presented in Fig.~\ref{fig_comp}. In particular, a feature vector $FV$ using CKSAAP for gap value of $k = 2$ can be obtained as 
\begin{align}
FV_{j=0} =& \left( \frac{F_{AA}}{N}, \frac{F_{AC}}{N}, \frac{F_{AD}}{N}, \dots, \frac{F_{YY}}{N} \right)_{400}, 
\\
FV_{j=1} =& \left( \frac{F_{AxA}}{N}, \frac{F_{AxC}}{N}, \frac{F_{AxD}}{N}, \dots, \frac{F_{YxY}}{N} \right)_{400}, 
\\
FV_{j=2} =& \left( \frac{F_{AxxA}}{N}, \frac{F_{AxxC}}{N}, \frac{F_{AxxD}}{N}, \dots, \frac{F_{YxxY}}{N} \right)_{400},
\\
\label{eqfv}
FV_{k=2} =& FV_{j=0} \mdoubleplus FV_{j=1} \mdoubleplus FV_{j=2} 	\in \mathbb{R}^{400\times(k+1)}=\mathbb{R}^{1200}.
\end{align}
where $N$ is the sequence length, $x$ is the gap or skipped residual, $F$ is the frequency count of amino acids pair, $\mdoubleplus$ denotes concatenation operation, and $A, C, D \dots Y$ are the standard symbols for amino acid. 

To generate a combined representation of both the E3 and its target proteins, we generated CKSAAP features from individual and concatenated protein sequences.  In total, we generated three representations of E3 and target proteins: (1) E3-sequence features (E3SF), (2) target-sequence features (TSF), and (3) concatenated sequence features called pair feature vector (PFV). The lengths of all three feature vectors are chosen to be the same, i.e., the same value of gap is used to extract CKSAAP features from all sequences. 
To avoid learning bias towards particular E3 or its target protein, we designed a modified auto-encoding scheme shown in Fig.~\ref{fig: ae}.  The objective is to learn a coupled latent space that can represent both E3 and its target with minimal complexity.  Towards this end, the auto-encoder model is designed to produce PFV using E3SF and TSF representations. This coupled learning helps in avoiding the input sequence bias towards a particular E3 or the target sequences, which is essential for a multi-input learning scenario. A detailed description of the proposed model is provided in the next section.


\subsection{Deep Latent Space Encoding}

One big advantage of CKSAAP encoding scheme is its high resolution since it can encode various combinations of amino-acid pairs at different levels of granularity. However, a high resolution comes at a price of noise and inflated feature space (i.e., high number of uninformative or redundant features).  In machine learning, it is well understood that for a robust classifier, a robust feature set that is minimally redundant and maximally relevant to the class label is important \cite{peng2005feature, naseem2017ecmsrc}. Therefore, to reduce dimensions of feature space, a variety of machine learning strategies has been proposed such as component analysis \cite{wang2018using}, information gain \cite{khan2016rafp}, and kernel or latent space encoding \cite{usman2020afp}.
    
Handcrafted features or manual curation of important features through feature engineering is useful when direct or simple (e.g., linear) relationship between the feature and the class label is required. In other words, manual curation is essential, when a linear and easily interpretable relationship between descriptor and target is needed to design experimentally verifiable studies. However, in most machine learning-based classification tasks, the descriptors are usually non-linearly related to class labels and show high inter-class variability. This poses a serious challenge for the section of useful feature space.
    
\begin{figure}[!htb]
	\begin{center}
		\centering \includegraphics[width=8cm]{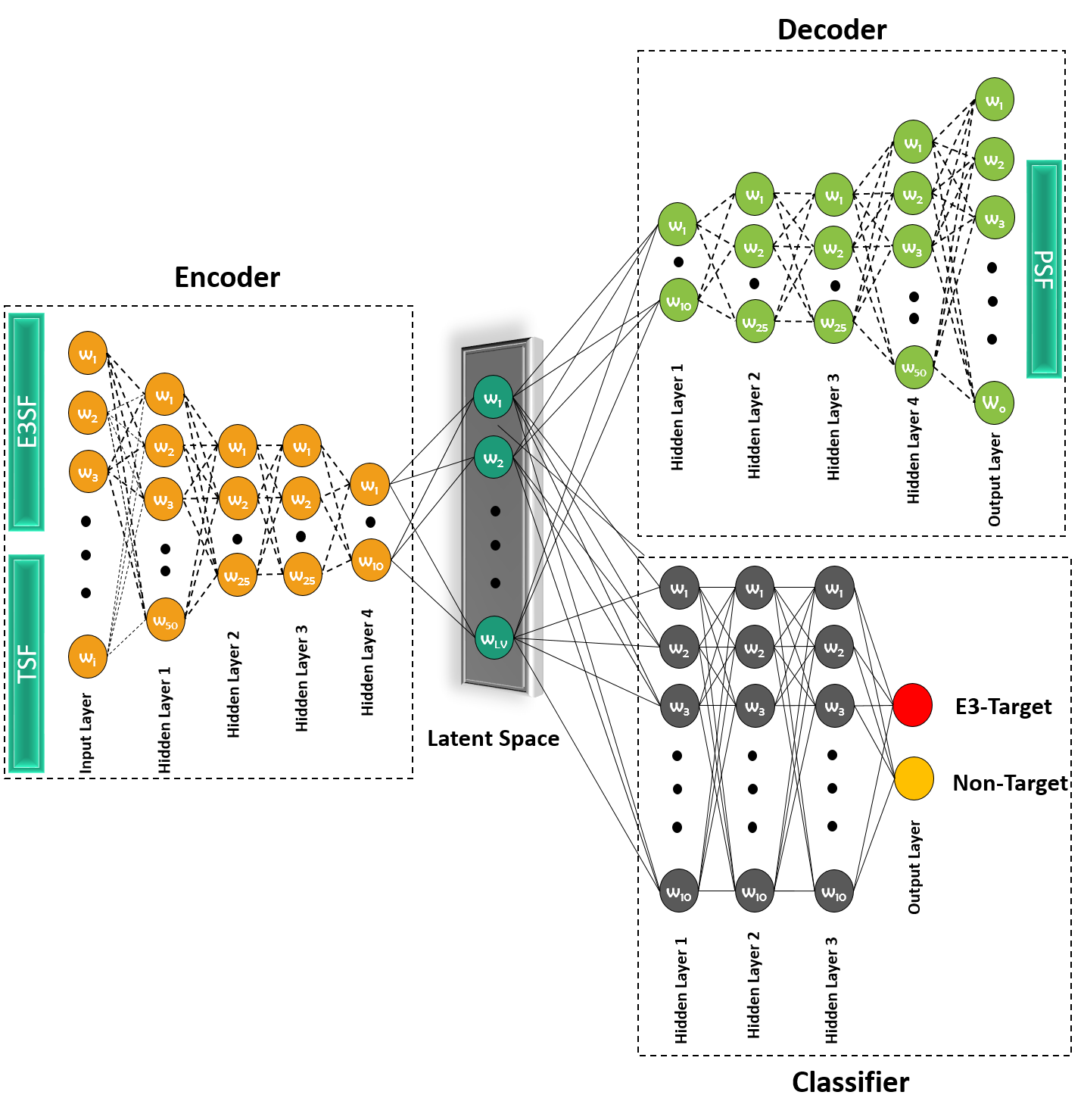}
	\end{center}
	\caption{Proposed latent-space encoding-based E3-target prediction model.}
	\label{fig: ae}
\end{figure}

Effective mapping of feature space to target label is a challenging task due to large feature space, non-linearity, and low inter-class and high intra-class variability. Towards this end,  an auto-encoder-based latent space encoding (AE-LSE) scheme is designed. In the AE-LSE scheme, the latent space of the auto-encoder is fed to the classifier, which imposes a constraint on auto-encoder to learn not only the descriptive features which are useful for minimizing reconstruction loss, but also the most discriminating features which are helpful for designing a power classifier model. Through this simultaneous optimization of auto-encoder and classifier models, a noise-free latent space representation can be learned. The architecture of the proposed AE-LSE based model is shown in Fig. \ref{fig: ae}.

\subsection{Training Configurations}
The model consists of five encoding and five decoding modules, each consists of batch-normalization, dropout, and ReLU activation function. The number of neurons in encoding block are $(50, 25, 25, 10)$.  The configuration of encoding and decoding blocks is chosen to be a mirror-symmetric as suggested in \cite{ye2018deep}, that is, the number of neurons in the  first hidden layer of the encoder is the same as the number of neurons in last hidden layer of the decoder. The same is true for second, third, and fourth hidden layers of auto-encoder, except input and output layers where the number of inputs (E3SF + TSF) is twice the number of the outputs (PSF). The classification module consists of three hidden layers each with $10$ neurons. For the output layer, the soft-max activation function is used and its size is set to be equal  to the number of class labels, i.e., two (E3-target, non-target). The model is implemented using Python on \emph{TensorFlow Keras}\cite{chollet2015keras} platform for variable size of latent space ($LV$), and gap values ($k$) specified in Section \ref{features}. For parameter optimization, two loss functions, i.e., the auto-encoder loss (mean-squared-error) and the classifier loss (binary-cross entropy), are minimized using \emph{Adam optimizer} on default learning rates for $1000$ epochs with early stopping tolerance of $25$ epochs.  The model used in this study is furnished online at (\href{https://github.com/psychemistz/E3targetPred}{https://github.com/psychemistz/E3targetPred}).

\subsection{Evaluation Parameters}

For a classification model, there are many widely used performance statistics. In particular,  Youden's index (YI) (or Youden's J statistics) \cite{youden1950index}, Matthew's Correlation Coefficient (MCC), and balanced accuracy (BACC) are considered to be the most comprehensive performance criteria.  The MCC ranges from $-1$ to $1$ with MCC = $1$ and MCC = $-1$ are the best and the worst predictions, respectively, and MCC = $0$ indicates the case of a random guess. For highly imbalanced test datasets, balanced accuracy and YI are interesting ways of summarizing the results of a diagnostic experiment.  The range of YI is from 0 to 1, where 0 indicates the worst performance while 1 shows perfect results with no false positives and no false negatives.  The proposed algorithm is extensively evaluated for these measures. For class-specific results, we also provide the true positive rate (sensitivity), and true negative rate (specificity).  Besides, we also provide reconstruction error statistics of the proposed AE-LSE scheme. The reconstruction error is calculated in the form of \emph{peak-signal-to-noise-ratio} (PSNR). The PSNR is calculated between the decoded output (estimated PSF) and the original PSF.

\section{Results}\label{sec:results}

\subsection{Ablation Study}
\begin{figure*}[!htb]
	\begin{center}
		\centering \includegraphics[width=12cm]{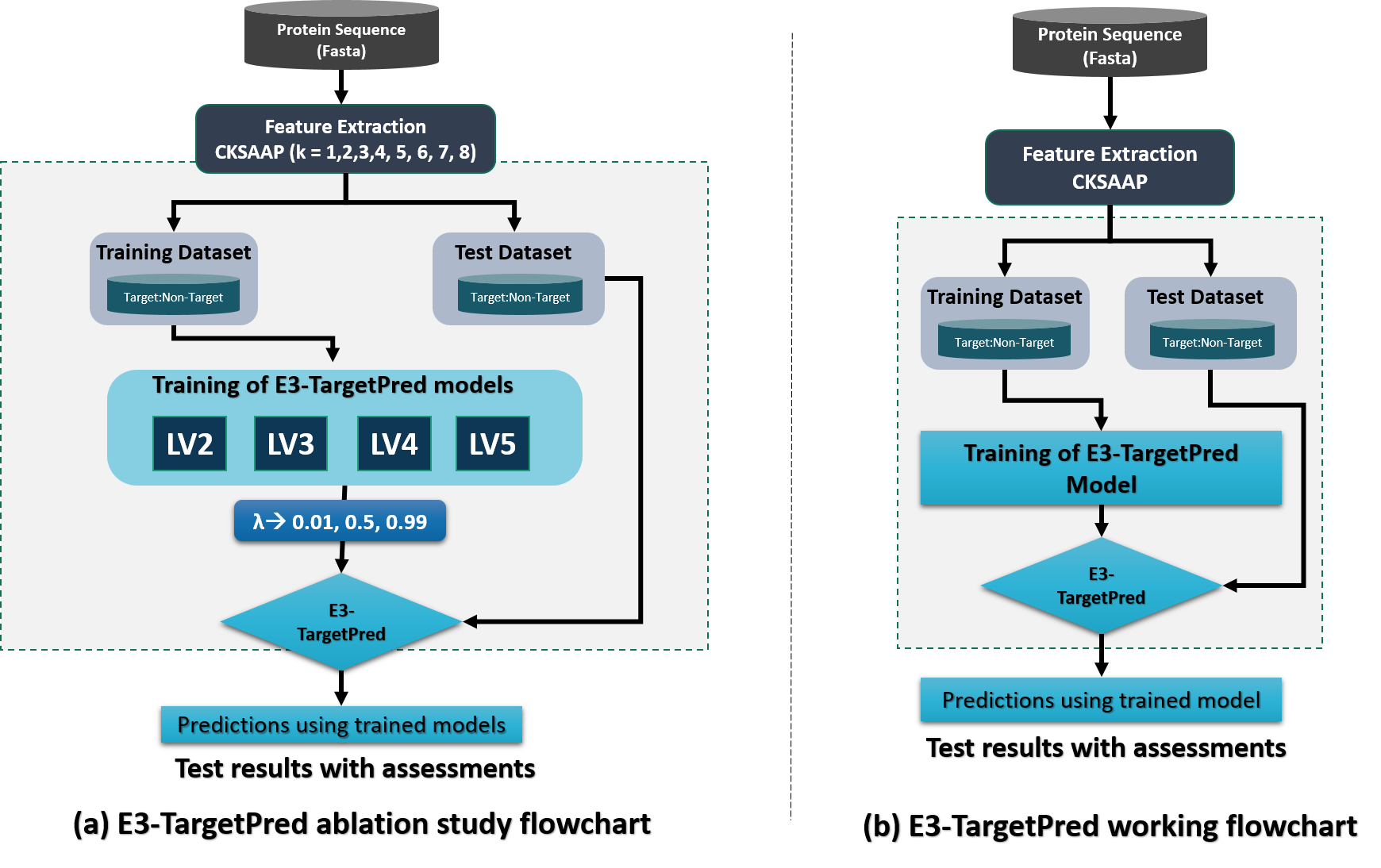}
	\end{center}
	\caption{Work-flow and ablation study diagram for E3-targetPred method.}
	\label{fig:workflow}
\end{figure*}

To investigate the effect of hyper-parameters on the performance of the proposed model, we designed an ablation study for the number of gaps ($k$) in CKSAAP  and the number of latent variables ($LV$) in AE-LSE model. We also compared three combinations of objective functions mixing weighting $\lambda $ (setting values $\lambda=0.01, 0.5, 0.99$) on the decoder loss, and ($1-\lambda$) on the classifier loss. A workflow diagram summarizing the ablation study is shown in Fig.~\ref{fig:workflow}(a). Test balanced accuracy for different combinations of $LV$, $k$, and $\lambda$ are shown in Table~\ref{tab:ablation_BACC}.  The objective of this thorough evaluation is to identify the best combination of all three parameters. In particular, we are interested in designing a powerful classifier with minimum complexity, specifically, with a latent space that we can use to easily visualize the clusters of E3-target and non-target pairs. The smallest value of the gap ensures minimum redundancy in the original feature space, which also reduces the computation overhead. 

In Table~\ref{tab:ablation_BACC}, for $20$ random trials mean and standard deviations in test balanced accuracy distributions for $108$ different combinations of $LV$, $k$, and $\lambda$ are presented. Overall, the proposed model shows stable performance for all given configurations, as it can be seen that the best BACC value of $76.1\%$ can be achieved with $\lambda=0.99$, $k=6$ and $LV=2$.   Therefore, the least complex best model in this study (i.e., $k=6$ and $LV=2$) is chosen for further testings. One big advantage of the $LV=2$ model is that its latent space can be visualized easily, which can help in finding the useful clusters for further studies.

\begin{table}[!htb]
\caption{Balanced accuracy results of ablation study on $\lambda$, Gap($k$), and $LV$ parameters.}
\begin{center}
\resizebox{0.45\textwidth}{!}{
\begin{tabular}{c c c c c c c}
\\ \hline
 $\lambda$ & ${\rm Gap}/LV$ & 2 & 3 & 4 & 5 \\ 
 \hline\hline
 & $0$ & $0.708\pm0.054$ & $0.720\pm0.026$ & $0.708\pm0.053$ & $0.718\pm0.026$ \\
 & $1$ & $0.720\pm0.023$ & $0.720\pm0.035$ & $0.712\pm0.056$ & $0.698\pm0.076$ \\
 & $2$ & $0.725\pm0.028$ & $0.684\pm0.069$ & $0.717\pm0.032$ & $0.708\pm0.051$ \\
 & $3$ & $0.730\pm0.036$ & $0.724\pm0.057$ & $0.743\pm0.026$ & $0.723\pm0.030$ \\
 $0.01$ & $4$ & $0.744\pm0.020$ & $0.737\pm0.018$ & $0.719\pm0.050$ & $0.712\pm0.053$ \\
 & $5$ & $0.730\pm0.026$ & $0.725\pm0.037$ & $0.726\pm0.055$ & $0.721\pm0.053$ \\
 & $6$ & $0.738\pm0.029$ & $0.719\pm0.037$ & $0.734\pm0.042$ & $0.730\pm0.036$ \\
 & $7$ & $0.735\pm0.031$ & $0.732\pm0.041$ & $0.740\pm0.025$ & $0.736\pm0.056$ \\
 & $8$ & $0.732\pm0.055$ & $0.743\pm0.022$ & $0.735\pm0.041$ & $0.733\pm0.033$ \\
  \hline\hline
 $\lambda$ & ${\rm Gap}/LV$ & 2 & 3 & 4 & 5\\ 
 \hline\hline
 & $0$ & $0.715\pm0.048$ & $0.727\pm0.025$ & $0.704\pm0.042$ & $0.698\pm0.066$ \\
 & $1$ & $0.694\pm0.064$ & $0.720\pm0.031$ & $0.700\pm0.074$ & $0.718\pm0.023$ \\
 & $2$ & $0.709\pm0.052$ & $0.729\pm0.041$ & $0.726\pm0.043$ & $0.721\pm0.048$ \\
 & $3$ & $0.697\pm0.081$ & $0.710\pm0.063$ & $0.730\pm0.062$ & $0.731\pm0.047$ \\
 $0.50$ & $4$ & $0.719\pm0.042$ & $0.723\pm0.048$ & $0.693\pm0.088$ & $0.721\pm0.056$ \\
 & $5$ & $0.706\pm0.058$ & $0.724\pm0.066$ & $0.719\pm0.062$ & $0.733\pm0.061$ \\
 & $6$ & $0.728\pm0.029$ & $0.727\pm0.063$ & $0.731\pm0.058$ & $0.755\pm0.021$ \\
 & $7$ & $0.744\pm0.035$ & $0.738\pm0.021$ & $0.741\pm0.015$ & $0.737\pm0.030$ \\
 & $8$ & $0.706\pm0.051$ & $0.723\pm0.057$ & $0.736\pm0.025$ & $0.720\pm0.067$ \\ 
 \hline\hline
 $\lambda$ & ${\rm Gap}/LV$ & 2 & 3 & 4 & 5\\ 
 \hline\hline
 & $0$ & $0.731\pm0.028$ & $0.722\pm0.039$ & $0.719\pm0.036$ & $0.740\pm0.034$ \\
 & $1$ & $0.741\pm0.027$ & $0.730\pm0.024$ & $0.726\pm0.031$ & $0.743\pm0.026$ \\
 & $2$ & $0.732\pm0.027$ & $0.740\pm0.034$ & $0.723\pm0.058$ & $0.745\pm0.019$ \\
 & $3$ & $0.740\pm0.023$ & $0.734\pm0.032$ & $0.747\pm0.031$ & $0.745\pm0.023$ \\
$0.99$ & $4$ & $0.738\pm0.031$ & $0.746\pm0.034$ & $0.741\pm0.029$ & $0.751\pm0.019$ \\
 & $5$ & $0.738\pm0.037$ & $0.739\pm0.061$ & $0.749\pm0.029$ & $0.747\pm0.028$ \\
 & $6$ & $0.761\pm0.021$ & $0.746\pm0.030$ & $0.752\pm0.023$ & $0.760\pm0.021$ \\
 & $7$ & $0.740\pm0.029$ & $0.755\pm0.025$ & $0.753\pm0.038$ & $0.752\pm0.030$ \\
 & $8$ & $0.747\pm0.028$ & $0.736\pm0.027$ & $0.749\pm0.030$ & $0.752\pm0.033$ \\ 
 \hline\hline
\end{tabular}
}
\end{center}
\label{tab:ablation_BACC}
\end{table}

\begin{figure}[!htb]
	\begin{center}
		\centering \includegraphics[width=8cm]{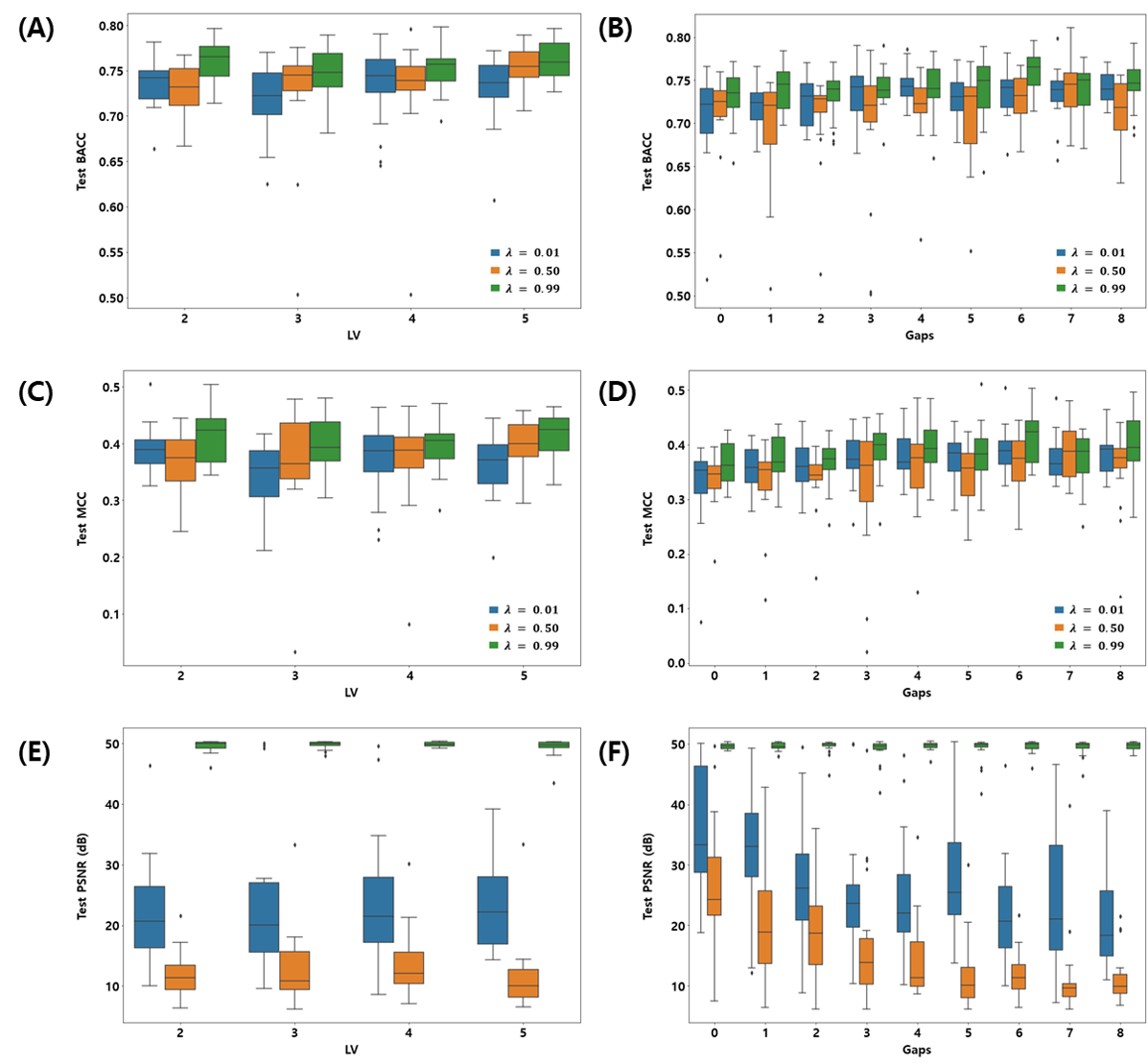}
	\end{center}
	\caption{Ablation study: Number of latent variables (LV) and Gaps in CKSAAP features are optimized within $4\times 9$ parameter space. Weights of the decoder  are chosen $0.01$ (blue), $0.5$ (yellow), and $0.99$ (green) with designated LV and Gaps.}
	\label{fig:ablation_4x9}
\end{figure}

The distributions of performance statistics for three combinations of $\lambda$ are delineated in Fig.~\ref{fig:ablation_4x9} to elucidate the effect of $\lambda$ on other performance measures.  In Figs.~\ref{fig:ablation_4x9}(A) and  \ref{fig:ablation_4x9}(B), distributions of BACC are shown for different LV and gap combinations, respectively. It is evident from the graphs that increasing decoder weight $\lambda$ improves the performance in all the cases.  A similar trend is observed in Figs.~\ref{fig:ablation_4x9}(C) and \ref{fig:ablation_4x9}(D) for MCC measure. However, it is more pronounced in Figs.~\ref{fig:ablation_4x9}(E) and \ref{fig:ablation_4x9}(F) for the decoder's reconstruction loss which is measured in terms of the PSNR. To better understand the origin of this performance gain due to increased $\lambda$, we analyze the latent space encoding schemes later on.

\subsection{Performance of the Best Model}

In Fig.~\ref{fig:workflow}(b), the workflow diagram of the proposed E3-Target study is shown. 
We chose the best model based on the result of the ablation study. As substantiated earlier, $LV = 2$ and $Gaps = 6$ performed the best among all the tested combinations. To evaluate the performance, we used BACC as the representative measure. The difference of performance between the validation and the test set is very small and the model successfully predicts unseen E3-target relationships in the testing phase (see Table~\ref{tab:bestmodel}). In particular, there are only $1.7\%$, $0.147$ units, and $0.033$ units drop compared to validation BACC, MCC, and YI, respectively, and no difference in the reconstruction error. One important aspect of the proposed model is that the difference between sensitivity and specificity is very low. This suggests that the model is well balanced and has a negligible bias toward a particular class type. This unbiased feature is important especially for a novel prediction case where no prior information about the given sample is known.
\begin{table}[!htb]
\caption{Performance statistics of (${\rm k} = 6, LV = 2$) model}
\begin{center}
\resizebox{0.45\textwidth}{!}{
\begin{tabular}{c c c c c c c}
\\ \hline
Measure & BACC & Sen & Spe & MCC & YI & PSNR (dB) \\ \hline\hline
$\mu_{\rm train}$ & 0.903  & 0.911  & 0.894  & 0.809  & 0.806  & 49.5  \\ \hline
$\sigma_{\rm train}$ & 0.036  & 0.065  & 0.051  & 0.070  & 0.071  & 0.938  \\ \hline
$\mu_{\rm valid}$ & 0.778  & 0.770  & 0.785  & 0.562  & 0.555  & 49.6  \\ \hline
$\sigma_{\rm valid}$ & 0.061  & 0.106  & 0.103  & 0.118  & 0.122  & 0.956  \\ \hline
$\mu_{\rm test}$ & 0.761  & 0.777  & 0.745  & 0.415  & 0.522  & 49.6  \\ \hline
$\sigma_{\rm test}$ & 0.021  & 0.071  & 0.068  & 0.049  & 0.042  & 1.019  \\ \hline
\end{tabular}
}
\bigskip
\end{center}
Mean $\mu_x$ and standard deviation $\sigma_x$ of measure in $x$ datasets ($20$ random trials). BACC: Balanced Accuracy, Sen: Sensitivity, Spe: Specificity, MCC: Mathew Correlation Coefficient, YI: Yoden Index, PSNR: Peak Signal to Noise Ratio $-\log_{10}({\rm MSE})$.
\label{tab:bestmodel}
\end{table}

\subsection{Comparison of Latent Space Encoding Schemes}

\begin{figure}[!htb]
	\begin{center}
		\centering \includegraphics[width=8cm]{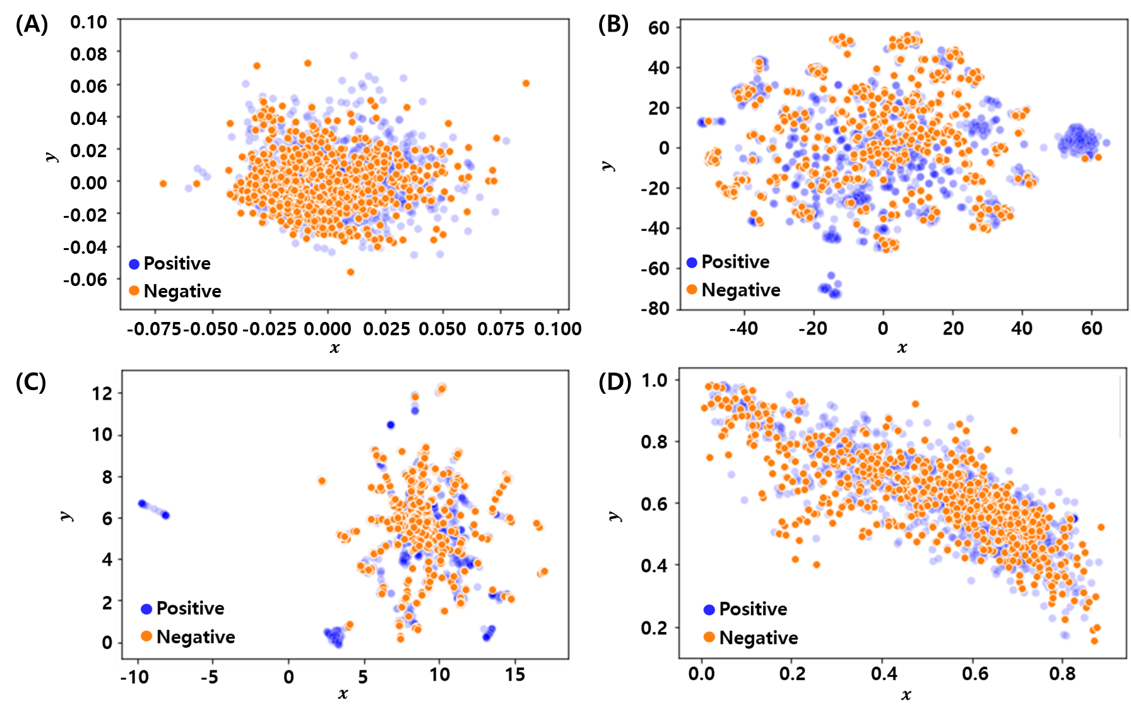}
	\end{center}
	\caption{Comparisons of feature space embeddings. (A) PCA, (B) t-SNE\cite{maaten2008visualizing}, (C) UMAP\cite{mcinnes2018umap}, and (D) Auto-encoder.}
	\label{fig:conventional_latentspace}
\end{figure}

To understand the origin of robust performance, we compared the latent space of proposed model (trained on different weight combinations of the decoder and classifier) with the standard dimension reduction methods. In Fig.~\ref{fig:conventional_latentspace}, PCA, t-SNE \cite{maaten2008visualizing},  UMAP \cite{mcinnes2018umap}, and  the auto-encoder are compared. It can be easily seen that the conventional methods failed to recover clusters of positive and negative E3-Targets from the original feature space. Especially in the case of the PCA and the auto-encoder, there is a complete overlap between positive and negative samples. In the case of the t-SNE and UMAP, there are a different number of small clusters but there are no clear distinguishing boundaries between two classes.

\begin{figure}[!htb]
	\begin{center}
		\centering \includegraphics[width=8cm]{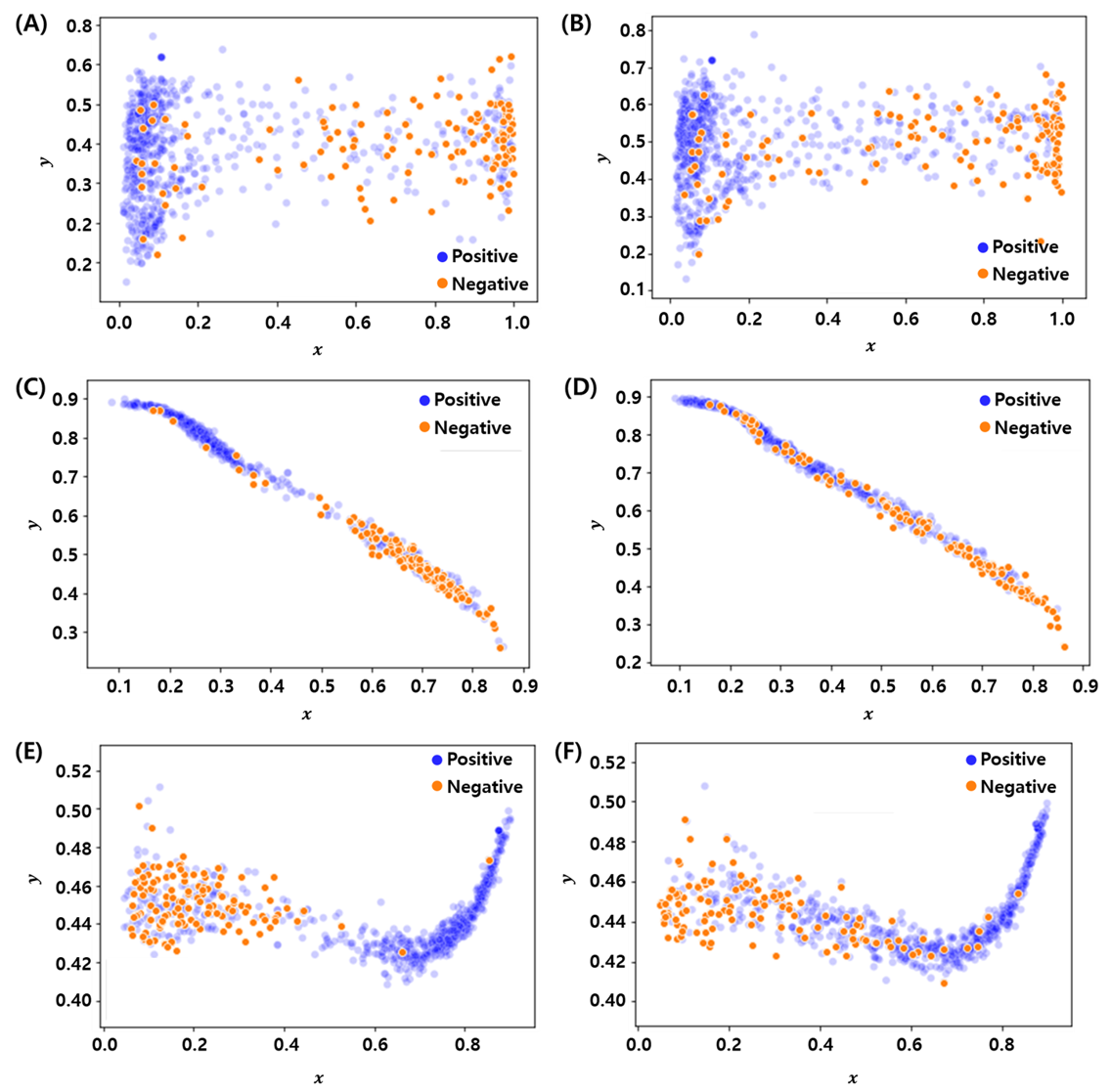}
	\end{center}
	\caption{Visualization of trained latent space (${\rm Gap}=6$, $LV=2$):
Positive (blue) and negative (yellow) E3-target relations in training set (left) and test dataset (right) mapped to trained latent space with $\lambda=0.99$, $\lambda=0.01$ and $\lambda=0.50$ (top to bottom). In $\lambda=0.99$ case, negative relations are clustered on the right side of latent space while positive relations are clustered on the left side of the space. The clustering of positive and negative data points is well maintained with the test dataset as well.}
	\label{fig:proposed_latentspace}
\end{figure}

In Fig.~\ref{fig:proposed_latentspace}, latent space learned by proposed method is visualized. Interestingly, both the positive and negative classes are exclusive in all cases. However, as expected with a large decoder weight, the model learns latent space with more variance. Indeed, since training attention is increased towards minimizing the reconstruction loss, the model has more freedom to represent individual samples with its variability. On the other hand, the attention of learning is towards minimizing the classification loss with a low value of the decoder weight. Therefore, the clusters are more compacted. Similar behavior is observed in the case of an equal weight of $\lambda$ but the problem is more challenging. Since both losses are equally weighted, it creates a tug of war between equally weighted objectives; the model is likely to learn from noisy latent space and performance of both the classifier and decoder drop (see, e.g., Fig.~\ref{fig:ablation_4x9}).

\subsection{Validation on Independent Test Dataset}

To evaluate the generalization power of proposed method, we tested the model on a completely independent test dataset. We used E3-substrate relation dataset curated by ESI network study \cite{chen2019multidimensional} and Negatome 2.0 \cite{blohm2014negatome}. We used catalogued E3 and target relations that are not included in our developmental dataset. We estimated the generalization performance of the model by dividing the four cases according to whether E3 or targets were included in the training data set (see Table~\ref{tab:ind_sample_stat}). 

The results are summarized in Table~\ref{tab:ind_peformance}. Overall, similar to the test dataset, the model show balanced total performance on independent dataset.  However as expected the performance in seen protein cases is higher compared to unseen cases. In particular, the balanced accuracy of the best model was around $0.814$ in seen E3/target case, but fell to $0.586$ in unseen E3/target case. In summary, although the performance was decreased in totally unseen E3/target cases, our model successfully retrieved about $72.9\%$ of known E3 and target relation in independent dataset.

\begin{table}[!htb]
\caption{Performance of the best model for independent datasets}
\begin{center}
\resizebox{0.45\textwidth}{!}{
\begin{tabular}{c c c c c c c}
\\ \hline
Dataset & BACC & Sen & Spe & MCC & YI & ACC  \\ \hline\hline 
Seen E3, Seen Target & 0.814  & 0.763  & 0.865  & 0.627  & 0.628  & 0.829   \\ \hline
Seen E3, UnSeen Target & 0.770  & 0.682  & 0.858  & 0.484  & 0.540  & 0.731  \\ \hline
UnSeen E3, Seen Target & 0.705  & 0.810  & 0.600  & 0.343  & 0.410  & 0.648  \\ \hline
UnSeen E3, Unseen Target & 0.586  & 0.600  & 0.572  & 0.121  & 0.172  & 0.576  \\ \Xhline{2\arrayrulewidth}
Total & 0.727  & 0.710  & 0.743  & 0.452  & 0.454  & 0.729  \\ \hline
\end{tabular}
}
\end{center}
\label{tab:ind_peformance}
\end{table}

\section{Discussion}\label{sec:discussion}

In this study, we proposed an E3-target prediction model based only on the sequence of two proteins. The proposed model was validated with an independent test dataset proposed by previous publications. We also verified that a trained E3-target prediction model can be generalized for unseen E3s as well as unseen targets. This property of our model is attractive as compared to the previous works. 

Several studies have been proposed to predict E3 and their substrate relationships. For example, Kai-Yao et al. \cite{huang2016new}, provided an interaction network viewer of E3 and target. Van-Nui Nguyen \emph{et al.} \cite{nguyen2016ubinet} proposed web resources for exploring ubiquitination in a network, called UbiNet. And Yang Li \emph{et al.} \cite{li2017integrated} proposed an integrated platform of E3-target relation called Ubibrowser. Yang Li \emph{et. al.} \cite{li2017integrated} provided the Naive-Bayes classifier for E3-target relation prediction based on multiple shreds of evidence such as homology, PPI, Gene ontology. Di Chen \emph{et al.} \cite{chen2019multidimensional} proposed multidimensional characterization of E3 and target interaction network by combining multiple sources. In particular, they trained a classifier of E3 and target relation based on expression datasets and network/pathway information derived features.

The aforementioned works provided a somewhat alternative view of E3-target relations but none of the proposed models was utilized for sequence-derived features alone. We hypothesized that if an E3 and its target proteins interact with each other there are classifiable features embedded in the composite sequence features space of E3 and targets.

Since our model utilizes only sequences of E3 and target, if the positive and negative relationships between E3 and target are not well characterized, the proposed model will be under-powered. Our LSE model has a unique advantage in this regard since it learns classifiable latent space while keeping characteristic of original features. In the proposed LSE, we can inspect whether predicted E3-target relations are closely distributed or not. Fig.~\ref{fig:proposed_latentspace} (A) shows the projection of data points in training dataset into trained latent space of $LV=2$. We can see that our model learned classifiable latent space which gives positive and negative class label clusters. Fig.~\ref{fig:proposed_latentspace} (B) shows the projection of data points in the test dataset. Although the classification boundary is not clear as in the training case, the test set also visualizes the distinction between positive and negative classes. Therefore, this latent space-based representation of E3-target relation enables us to calculate confidence or refine the threshold of E3-target relation to reduce potential false positives. 

\section{Conclusion} \label{sec:conclusion}

In this study, for the first time, we proposed a sequence-based E3-target relation prediction model called E3targetPred. Based on the comprehensive ablation study, we characterized the optimal number of gaps and latent variables to be utilized in our model. Besides, we compared the performance of conventional latent space encoding schemes and substantiated that the proposed model provides separable clusters in latent space.  E3-targetPred has a unique advantage compared to conventional models mainly due to the latent space learning property of the model. Since it learns non-linear embedding of the features while keeping properties of the original feature space, we can further filter out suspicious data points based on the distance between a data point and the class center. This further increases the confidence of the classifier for noisy annotations. The code and dataset utilized in this work are provided at GitHub page of the author (\href{https://github.com/psychemistz/E3targetPred}{https://github.com/psychemistz/E3targetPred})





\end{document}